\documentstyle[aps,prd,epsfig,preprint]{revtex}
\def\fun#1#2{\lower3.6pt
\vbox{\baselineskip0pt\lineskip.9pt
\ialign{$\mathsurround=0pt#1\hfill##\hfil$
\crcr#2\crcr\sim\crcr}}}
\begin{document}
\vspace{0.5in}
\title{\vskip-2.5truecm{\hfill \baselineskip 14pt 
{\hfill {{\small \hfill UT-STPD-4/00}}}\\
{{\small \hfill BA-00-29}}
\vskip .1truecm} 
\vspace{1.0cm}
\vskip 0.1truecm{\bf Monopoles, Axions and Intermediate 
Mass Dark Matter}}
\vspace{1cm}
\author{{G. Lazarides}$^{(1)}$\thanks{lazaride@eng.auth.gr} 
{and Q. Shafi}$^{(2)}$\thanks{shafi@bartol.udel.edu}} 
\vspace{1.0cm} 
\address{$^{(1)}${\it Physics Division, School of Technology, 
Aristotle University of Thessaloniki,\\ 
Thessaloniki 540 06, Greece.}}
\address{$^{(2)}${\it Bartol Research Institute, University
of Delaware, Newark, DE 19716, USA.}}
\maketitle

\vspace{1.5cm}

\begin{abstract}
\baselineskip 12pt

\par
We present a solution to the cosmological problem 
encountered in (supersymmetric) grand unified theories 
due to copious monopole production at the end of hybrid 
inflation. By employing thermal inflation ``driven'' by the 
$U(1)$ axion symmetry, the superheavy monopole flux can be 
naturally suppressed to values that should be accessible to 
dedicated large scale experiments. The $U(1)$ axion 
symmetry also helps generate the right magnitude for the 
$\mu$ term of the minimal supersymmetric standard model. 
An important by-product is the predicted existence of stable 
or very long-living fermions possessing intermediate scale 
masses of order $10^{12}~{\rm{GeV}}$. Their presence is 
required for implementing thermal inflation, and their 
stability is due to a $Z_2$ symmetry. They may constitute 
a sizable fraction of cold dark matter, and possibly help 
explain the ultra-high energy cosmic ray events. The rest 
of cold dark matter may consist of axions. Although our 
discussion is carried out within the framework of 
supersymmetric $SU(4)_c\times SU(2)_L\times SU(2)_R~$, 
it can be extended to other grand unified gauge groups such 
as $SU(3)_c\times SU(3)_L\times SU(3)_R$ or $SO(10)$.
\end{abstract}

\thispagestyle{empty}
\newpage
\pagestyle{plain}
\setcounter{page}{1}
\baselineskip 20pt

\par
The great advantage of hybrid inflation \cite{linde} is 
that, in contrast to previous inflationary schemes, 
it can reproduce the observed temperature fluctuations of 
the cosmic microwave background radiation with natural 
values of the relevant coupling constant. Moreover, this 
inflationary scenario is almost automatically realized 
\cite{lyth,dss} in supersymmetric (SUSY) grand unified 
theories (GUTs). However, in trying to apply it to GUTs 
which predict the existence of magnetic monopoles, a 
cosmological disaster is encountered. Hybrid inflation 
is terminated abruptly when the system reaches an 
instability point on the inflationary trajectory and is 
followed by a `waterfall' regime during which the 
spontaneous breaking of the GUT gauge symmetry takes 
place. The appropriate Higgs fields develop their 
vacuum expectation values (vevs) starting from zero and 
they can end up at any point of the vacuum manifold with 
equal probability. As a consequence, monopoles are 
copiously produced \cite{smooth} by the Kibble 
mechanism \cite{kibble} leading to a cosmological 
catastrophe.

\par
Possible solutions to this monopole problem have been 
proposed \cite{smooth,jean}. They rely on introducing 
the leading non-renormalizable term in the standard 
superpotential \cite{lyth} for hybrid inflation. (For 
different resolutions of the problem see 
Ref.\cite{dvali}.) In Ref.\cite{smooth}, the trilinear 
coupling of this standard superpotential was eliminated 
by a discrete symmetry and was replaced by the leading 
non-renormalizable term. The system, from the beginning 
of inflation, follows a particular valley and ends up at 
a particular point of the vacuum manifold. Thus, no 
monopoles can be produced. The inflationary trajectory 
possesses a classical inclination driving the inflaton 
towards the SUSY vacua and the termination of inflation 
is smooth. In Ref.\cite{jean}, both the trilinear and 
the leading non-renormalizable couplings were kept 
revealing a quite different picture. The inflationary 
trajectory is classically flat and, thus, radiative 
corrections \cite{dss} are needed for driving the 
inflaton. The termination of inflation is abrupt. 
Nevertheless, there is no monopole production since the 
GUT gauge symmetry is already broken during inflation. 
Both models predict complete absence 
of monopoles which may be disappointing for the 
experimenters.

\par
In this letter, we propose an alternative solution to the 
monopole problem of hybrid inflation which may yield a 
measurable monopole flux in our galaxy. The idea is to 
keep the original SUSY hybrid inflationary scenario 
unaltered and try to dilute the monopoles by invoking a 
subsequent thermal inflation \cite{thermal1,thermal2} 
associated with an intermediate mass scale. (The main 
ideas underlying thermal inflation have been presented 
in Ref.\cite{thermal1}. The term thermal inflation was 
coined in Ref.\cite{thermal2}, which further elaborated 
on the scheme.) It is then natural to 
identify this scale with the one at which the Peccei-Quinn 
(PQ) symmetry \cite{pq} is broken and also use \cite{rsym} 
it for generating the $\mu$ term of the minimal 
supersymmetric standard model (MSSM). Although of much wider 
applicability, our mechanism is presented in the context of 
the SUSY Pati-Salam (PS) model \cite{ps}, which is one the 
simplest unified schemes possessing \cite{magg} monopoles. 

\par
Our mechanism has an interesting by-product. In trying 
to make thermal inflation possible, we are led to the 
introduction of a number of superfields coupled to the 
field which breaks spontaneously the PQ symmetry. These 
fields possess intermediate scale masses, with the 
lightest ones being either stable or very long-living 
as a consequence of a $Z_2$ symmetry. Their fermionic 
components may constitute a sizable fraction of the cold 
dark matter in the universe, and possibly help explain 
\cite{sdm} the ultra-high energy cosmic rays \cite{uhecr}. 
Axions, of course, also may contribute to the cold dark 
matter fraction.

\par
We consider the SUSY PS model \cite{ps} with gauge group  
$G_{PS}=SU(4)_c\times SU(2)_L\times SU(2)_R~$. 
The left-handed quark and lepton superfields are accommodated 
in the representations $F_i=(4,2,1)$, $F^c_i=(\bar{4},1,2)$, 
where $i=1,2,3$ is the generation index. The two 
electroweak Higgs superfields belong to the representation 
$h=(1,2,2)$. The PS gauge group can be spontaneously broken 
to the standard model gauge group by a conjugate pair of Higgs 
superfields $H^c=(\bar{4},1,2)$, $\bar{H}^c=(4,1,2)$ 
acquiring non-vanishing vevs along their right-handed neutrino 
directions. This can be achieved by introducing a gauge singlet 
superfield $S$ with two (renormalizable) superpotential terms: 
a term linear in $S$ and a trilinear coupling of $S$ to 
$H^c$, $\bar{H}^c$. The resulting scalar potential 
automatically possesses an in-built (classically) flat 
direction along which hybrid inflation can take place 
\cite{lyth} with the system driven by an inclination from 
one-loop radiative corrections \cite{dss}. $G_{PS}$ is 
restored along the inflationary trajectory and breaks 
spontaneously only at the end of inflation when the system 
falls towards the SUSY vacua. This transition leads 
\cite{smooth} to a cosmologically catastrophic copious 
production of doubly charged monopoles \cite{magg}. The 
monopoles could be diluted to an acceptable level if the 
primordial hybrid inflation is followed by thermal inflation 
\cite{thermal1,thermal2}. This inflation, associated with an 
intermediate mass scale, is terminated at cosmic temperatures 
of the order of the electroweak scale and generates only a 
moderate number of e-foldings.

\par
Thermal inflation could be ``driven'' by the PQ symmetry 
which is spontaneously broken at an intermediate scale. 
Moreover, in Ref.\cite{rsym}, we showed that the PQ 
symmetry, which solves the strong CP problem, can also be used 
to generate the $\mu$ term of MSSM with the desired magnitude. 
More specifically, we introduced a pair of gauge singlet 
superfields $N$, $\bar{N}$ with non-zero PQ charges and a 
(non-renormalizable) superpotential coupling $N^2\bar{N}^2$. 
The resulting scalar potential of these fields, after soft SUSY 
breaking, possesses a non-trivial minimum which, under certain 
circumstances, is the absolute minimum, with $N$, $\bar{N}$ 
acquiring intermediate scale vevs, thereby breaking the PQ 
symmetry. The $\mu$ term is then generated via the 
superpotential coupling $N^2h^2$. The 
trivial extremum (at $N=\bar{N}=0$) turns out to be a local 
minimum separated from the PQ minimum by a sizable potential 
barrier. This situation persists at all cosmic temperatures 
after the primordial reheating which follows hybrid inflation, 
as has been shown \cite{jean} by including the one-loop 
temperature corrections to the potential. Thus, a 
successful transition from the trivial to the PQ minimum 
cannot be realized. We had to assume \cite{jean} that the 
system, after the primordial reheating, already emerges in 
the PQ vacuum. In other words, there is neither PQ transition 
nor thermal inflation in this case.

\par
In order to make thermal inflation possible, we must turn the
trivial local minimum of the zero-temperature potential into 
a saddle point. More specifically, the positive soft 
${\rm{mass}}^2$ term of $N$ should become negative. This 
can be achieved radiatively and requires strong couplings of 
$N$ to a number of superfields. To this end, we introduce 
$n$ superfields $D_a$ ($a=1,2,...,n$) belonging to the 
representation $(6,1,1)$ of $G_{PS}$ with superpotential 
couplings $NDD$. However, these color (anti)triplets acquire 
intermediate scale masses after the PQ breaking, which could 
prevent the unification of the MSSM gauge coupling constants. 
To restore gauge unification, we include an equal number of 
superfields $H_a$ belonging to the $(1,2,2)$ representation 
with superpotential couplings $NHH$. Note that the negative 
${\rm{mass}}^2$ term of $N$, which is successfully generated 
by invoking these extra superfields, gives rise to an 
additional problem. All the higher order terms in the scalar 
potential, which are derived from $N^2\bar{N}^2$ after soft 
SUSY breaking, involve both $N$, $\bar{N}$. Thus, 
these terms vanish along the $N$ direction, and the potential 
becomes unbounded below due to the negative ${\rm{mass}}^2$ 
term of $N$. Fortunately, there is a simple way out from this 
``runaway" problem. Replacing $N^2\bar{N}^2$ by 
$N^3\bar{N}$, we generate a $|N|^6$ term in the potential 
which prevents the runaway behavior along the $N$ direction.   
        
\par
We still need to include some extra couplings and superfields 
to obtain a phenomenologically viable scheme. In particular, 
we must introduce quartic superpotential couplings of 
$\bar{H}^c$ to $F^c_i$. These couplings generate 
intermediate scale masses for the right-handed 
neutrinos and, thus, seesaw masses for the light neutrinos. 
The inflaton then decays into right-handed neutrinos via the 
same couplings. Finally, in order to give superheavy masses 
to the down quark type components of $H^c$, $\bar{H}^c$, 
we include \cite{leont} an $SU(4)_c$ 6-plet superfield 
$G=(6,1,1)$ with superpotential couplings $GH^cH^c$, 
$G\bar{H}^c\bar{H}^c$. 

\par
The superpotential of the model, which incorporates all the
above couplings, is
\begin{eqnarray}   
W=\kappa S(\bar{H}^c H^c-M^2)
+\lambda_1 \frac{N^2 h^2}{m_P}
+\lambda_2 \frac{N^3 \bar{N}}{m_P}
+\alpha_{a}ND_aD_a+\beta_{a}NH_aH_a
\nonumber \\
+\lambda_{ij} F^c_i F_j h
+\gamma_{ij} \frac{\bar{H}^c \bar{H}^c}{m_P} F_i^cF_j^c 
+ a G H^c H^c + b G \bar{H}^c \bar{H}^c,~~~~~~~~~~~~
\label{eq:superpot}
\end{eqnarray}
where $M$ is a superheavy scale and 
$m_P=M_P/\sqrt{8\pi}\approx 2.44\times 10^{18}~{\rm{GeV}}$ 
is the reduced Planck mass. Here, we chose a basis in the $D_a$, 
$H_a$ space where the coupling constant matrices $\alpha$ 
and $\beta$ are diagonal. Assuming that, at a more fundamental 
level, the $D$'s and $H$'s originate from $SO(10)$ 10-plets, 
we can obtain the restriction $\alpha_{a}=\beta_{a}$ 
($a=1,2,...,n$). Note that $M$, $\kappa$, $\lambda_{1,2}$, 
$\alpha_a$, $\beta_a$, $a$ and $b$ can be made positive by 
field redefinitions. 

\par
In addition to $G_{PS}$, the superpotential in 
Eq.(\ref{eq:superpot}) possesses two continuous global 
(anomalous) symmetries, namely a R symmetry $U(1)_R$ and a 
PQ symmetry $U(1)_{PQ}~$. The R and PQ charges of the 
superfields are assigned as follows:
\begin{eqnarray}
R:H^c(0), \bar{H}^c(0), S(4), G(4), D(1), H(1), F(2), 
F^c(2), N(2), \bar{N}(-2), h(0);~~ 
\nonumber \\ 
PQ:H^c(0), \bar{H}^c(0), S(0), G(0), D(1), H(1), F(-2), 
F^c(0), N(-2), \bar{N}(6), h(2).
\label{eq:charges}
\end{eqnarray}
Note that the R charge of $W$ is 4. Although it is not necessary, 
we also impose, for simplicity, a discrete $Z_2^{mp}$ symmetry 
(``matter parity"), under which $F$, $F^c$ change sign. 
Additional superpotential terms allowed by the symmetries of the 
model are
\begin{equation}
FFH^cH^cN\bar{N},~FFH^cH^ch^2,~FF\bar{H}^c\bar{H}^cN\bar{N},
~FF\bar{H}^c\bar{H}^ch^2,~F^cF^cH^cH^c, 
\label{eq:terms}
\end{equation}
modulo arbitrary multiplications by non-negative powers of the
combinations $H^c\bar{H}^c$, $(H^c)^4$, $(\bar{H}^c)^4$ 
(this applies to the terms in Eq.(\ref{eq:superpot}) too). 
Note that, without the $Z_2^{mp}$ symmetry, the coupling 
$DHF\bar{H}^c$ would also be present in Eq.(\ref{eq:terms}).

\par
Instanton and soft SUSY breaking effects explicitly break
$U(1)_R\times U(1)_{PQ}$ to a discrete subgroup. It is then 
important to ensure that this subgroup is not spontaneously
broken by the vevs of $N$, $\bar{N}$ since otherwise 
cosmologically disastrous domain walls will be produced in the 
PQ transition. This requirement implies that the number $n$ of 
$D$'s and $H$'s must be 5 or 7. Moreover, in both these cases, 
the subgroup of $U(1)_R\times U(1)_{PQ}$ left 
unbroken by instantons and SUSY breaking coincides with the 
one left unbroken by $\langle N\rangle$, 
$\langle\bar{N}\rangle$, and is a $Z_2\times Z_2$ 
generated by $(e^{i\pi/2},e^{i\pi/2})$ and 
$(1,e^{i\pi})$. (Note that the element 
$(e^{i\pi},e^{i\pi})$ is equivalent to the identity 
element since it leaves unaltered all the superfields of the 
theory.) Combining appropriately these $Z_2$'s with $Z_2^{mp}$
and the $Z_2$ center of $SU(2)_L$, we obtain two equivalent
$Z_2$'s under which the $D$'s or the $H$'s change sign. It is 
interesting to note that, even if only $N$ develops a non-zero 
vev (see below), absence of domain walls still implies 
$n=5~{\rm{or}}~7$. The soft SUSY breaking terms respect the 
symmetry $ Z_2\times U(1)_{PQ}$, where the non-anomalous $Z_2$ 
is generated by $(e^{i\pi/2},e^{i\pi/2})$. It is then obvious 
that further breaking of $U(1)_{PQ}$ to $Z_2$ by 
the vev of $N$ can solve the strong CP problem. Thus, we have the 
option to keep $\langle\bar{N}\rangle=0$. The symmetries which 
survive after instanton effects, in this case, are the same as in 
the $\langle\bar{N}\rangle\neq 0$ case. 

\par
Let us note that baryon and lepton number violations arise from 
the last three superpotential terms in Eq.(\ref{eq:terms}), 
the last two couplings in Eq.(\ref{eq:superpot}) and the 
combinations $(H^c)^4$, $(\bar{H}^c)^4$, in complete analogy 
with Ref.\cite{jean}. The proton is practically stable.

\par
The part of the tree-level scalar potential which is relevant 
for the PQ (and R symmetry) breaking can be derived from the 
superpotential term $N^3\bar{N}$ and, after soft SUSY 
breaking mediated by minimal supergravity, is given by (compare 
with Ref.\cite{rsym})
\begin{equation}
V_{PQ}=m_{3/2}^2\left(\vert N\vert^2+\vert \bar{N}\vert^2
+\lambda_2^2\frac{\vert N\vert^4}{(m_{3/2}m_P)^2}
(\vert N\vert^2+9\vert\bar{N}\vert^2)-
\vert A\vert\lambda_2 
\frac{\vert N\vert\vert\bar{N}\vert}
{m_{3/2}m_P}\vert N\vert^2\right),
\label{eq:pqpot}
\end{equation}
where $m_{3/2}$ is the gravitino mass and $A$ is the 
dimensionless coefficient of the soft SUSY breaking term which 
corresponds to the superpotential coupling $N^3\bar{N}$. Here, 
the phases $\varphi$, $\theta$ and $\bar{\theta}$ of $A$, 
$N$ and $\bar{N}$ are taken to satisfy the relation 
$\varphi+3\theta+\bar{\theta}=\pi$ which minimizes the 
potential for given values of $\vert N\vert$, 
$\vert\bar{N}\vert$.

\par
As a consequence of the couplings of $N$ to the $D$'s and $H$'s, 
its ${\rm{mass}}^2$ in Eq.(\ref{eq:pqpot}) can easily turn 
radiatively to negative values at lower energy scales (see also 
Ref.\cite{choi}). To see this, we consider the one-loop 
renormalization group equations for the soft masses $m_N$, $m_D$ 
and $m_H$ of the scalar fields $N$, $D$ and $H$ (see 
Ref.\cite{mangano}):
\begin{eqnarray}
\mu\frac{dm_N^2}{d\mu}=3nY_D(2m_D^2+m_N^2)
+2nY_H(2m_H^2+m_N^2)~,~~~
\nonumber \\
\mu\frac{dm_D^2}{d\mu}=Y_D(2m_D^2+m_N^2)~,
~\mu\frac{dm_H^2}{d\mu}=Y_H(2m_H^2+m_N^2)~,
\label{eq:rge}
\end{eqnarray}
where $\mu$ is the running energy scale,
$Y_D=\alpha^2/2\pi^2$, $Y_H=\beta^2/2\pi^2$, with 
$\alpha=\alpha_a$, $\beta=\beta_a$ ($a=1,2,...,n$). 
Here we take universal ``asymptotic" soft scalar masses 
($m_N=m_D=m_H=m_{3/2}$ at the GUT scale) and assume, 
for simplicity, that all the $\alpha$'s ($\beta$'s) are 
equal and, thus, all the $D$'s ($H$'s) have the same soft 
mass at all scales. This system of equations possesses 
\cite{mangano} a non-trivial fixed point given by
\begin{equation}
m_N^2=-2m_D^2=-2m_H^2=-\frac{2(5n-1)}{5n+2}m_{3/2}^2~,
\label{eq:fixed}
\end{equation}
and admits the solution
\begin{eqnarray}
\frac{m_N^2}{m_{3/2}^2}=-\frac{2(5n-1)}{5n+2}+
\frac{15n}{5n+2}\left(\frac{\mu}{M_G}\right)^{(5n+2)Y},
~~~
\nonumber \\
\frac{m_D^2}{m_{3/2}^2}=\frac{m_H^2}{m_{3/2}^2}=
\frac{5n-1}{5n+2}+\frac{3}{5n+2}
\left(\frac{\mu}{M_G}\right)^{(5n+2)Y}.
\label{eq:solution}
\end{eqnarray}
For simplicity, we further assumed that $\alpha=\beta$ (and, 
thus, $Y_D=Y_H=Y$), and we ignored the running of these  
coupling constants. Taking strong Yukawa couplings 
$\alpha=\beta=1$, we see that, already at 
$\mu/M_G\sim 10^{-2}$, the second terms in the right hand sides 
of the equalities in Eq.(\ref{eq:solution}) are much smaller than 
$1\%$ of the first terms. Thus, after the primordial reheating, 
the soft masses can be taken equal to their fixed point values.

\par
The radiatively improved zero-temperature scalar potential is 
given by Eq.(\ref{eq:pqpot}) with the $\vert N\vert^2$ term
acquiring an extra negative factor $-2(5n-1)/(5n+2)$. The trivial 
(local) minimum of this potential (at $N=\bar{N}=0$) then becomes 
a saddle point and the absolute minimum necessarily lies at a 
non-zero value of $N$. It should be noted, however, that $\bar{N}$ 
is also non-zero at the absolute minimum only if $A\neq 0$. To 
simplify the presentation, we choose $A=0$. The absolute minimum 
of the potential then lies at
\begin{equation}
\vert\langle N\rangle\vert\equiv \frac{f_a}{\sqrt{2}}=
\left(\frac{10n-2}{15n+6}\right)^{1/4}
\frac{(m_{3/2}m_P)^{1/2}}{\lambda_2^{1/2}}~,~
\vert\langle\bar{N}\rangle\vert=0.
\label{eq:minimum}
\end{equation}
The constant energy density carried by the trivial vacuum is given 
by
\begin{equation}
V_0=2\left(\frac{10n-2}{15n+6}\right)^{3/2}
\frac{m_{3/2}^3m_P}{\lambda_2},
\label{eq:energy}
\end{equation}
and is responsible for driving thermal inflation.

\par
The one-loop temperature corrections to the PQ potential can be
calculated by employing the formalism of Ref.\cite{temperature}.
The absolute minimum of the resulting temperature-dependent 
effective potential at high cosmic temperatures $T$ lies at 
$N=\bar{N}=0$. So, after the primordial reheating following 
hybrid inflation, the system emerges in the trivial vacuum. The 
temperature correction to the ${\rm{mass}}^2$ term of the 
field $N$ is
\begin{equation}
\delta V_T\approx nT^2(3\alpha^2+2\beta^2)\vert N\vert^2,
\label{eq:temperature}
\end{equation}
for $\vert N\vert\ll T$. Here, we considered only the main 
contributions which originate from loops with the superfields 
$D$ or $H$ circulating. As $T$ approaches the critical 
temperature 
\begin{equation}
T_c\approx
\sqrt{\frac{10n-2}{(5n^2+2n)(3\alpha^2+2\beta^2)}}
~m_{3/2}~,
\label{eq:critical}
\end{equation} 
the potential barrier separating the trivial and PQ vacua 
becomes vanishingly small and the PQ transition takes place 
close to $T_c$ \cite{thermal1}. For a period 
preceding this transition, the energy density of the trivial 
vacuum dominates over the radiation energy density and the 
universe undergoes a mild inflationary phase known as thermal 
inflation \cite{thermal1,thermal2}. This is terminated at 
the PQ transition where the system enters into an oscillatory 
phase about the PQ minimum. The coherently oscillating field 
$N$ (thermal inflaton) with mass 
$m_{infl}=2[(5n-1)/(5n+2)]^{1/2}m_{3/2}$ eventually 
decays, via the superpotential coupling $N^2h^2$, to a pair 
of Higgsinos thereby reheating the universe. Of course, this 
decay is possible only if the Higgsino mass 
$\mu=\lambda_1f_a^2/m_P$ 
is smaller than $m_{infl}/2$. The decay width is 
$\Gamma\approx(2\lambda_1^2f_a^2/\pi m_P^2)
(1-\epsilon^2)^{1/2}m_{infl}$ where 
$\epsilon=2\mu/m_{infl}$ ($0<\epsilon<1$). The maximal 
reheat temperature is achieved at $\epsilon^2=2/3$, which 
maximizes $\Gamma$, and is given by
\begin{equation}
T_r\approx\frac{30^{1/4}\lambda_1f_am_{infl}^{1/2}}
{\pi g_*^{1/4}(T_r)m_P^{1/2}}~,
\label{eq:reheat}
\end{equation}
where $g_*(T_r)\approx 89.75$ is the effective number 
of degrees of freedom at the reheat temperature 
$T_r\sim {\rm{GeV}}$ (see below).

\par
The relative monopole number density $n_M/s$ ($n_M$ is 
the number density of monopoles and $s$ the entropy 
density) remains \cite{preskill} essentially constant for 
temperatures between the primordial reheat temperature 
$T_R\stackrel{_{<}}{_{\sim }}10^{12}~{\rm{GeV}}$ 
(see below) and the critical temperature $T_c~$, where 
the vacuum energy density $V_0$ is transferred to the 
oscillating inflaton field. The initial number density of 
thermal inflatons is  
$n_{infl}\approx V_0/m_{infl}~$. So, at $T_c~$,
\begin{equation} 
\frac{n_M}{n_{infl}}\approx\frac{n_M}{s}~
\frac{2\pi^2g_*(T_c)T_c^3m_{infl}}{45V_0}~,
\label{eq:monopole}
\end{equation}
where $g_*(T_c)\approx 105.75+35n/2$, for 
$T_c\sim 70~{\rm{GeV}}$ (see below), is the effective 
number of degrees of freedom just before the PQ transition. 
Note that, for temperatures between $T_R$ and $T_c$, the 
fermionic components of the $D$'s and $H$'s are massless. 
The ratio in Eq.(\ref{eq:monopole}) remains \cite{mono} 
practically unaltered until $T_r~$, where 
$n_M/s=(n_M/n_{infl})(n_{infl}/s)$ 
with $n_{infl}/s\approx 3T_r/4m_{infl}$ in the 
instantaneous inflaton decay approximation. Combining this 
with Eq.(\ref{eq:monopole}), one obtains the dilution 
factor of the relative monopole number density during the 
PQ transition and subsequent reheating:
\begin{equation}
\left(\frac{n_M}{s}\right)(T_r)\approx
\left(\frac{n_M}{s}\right)(T_R)~
\frac{\pi^2g_*(T_c)T_c^3T_r}{30V_0}~\cdot
\label{eq:dilution}
\end{equation}

\par
At the primordial reheating, we have 
$n_M/s\approx(n_M/n_{INFL})(3T_R/4m_{INFL})$, where 
$m_{INFL}=\sqrt{2}\kappa M$ ($n_{INFL}$) is \cite{lss} 
the inflaton mass (number density) for hybrid inflation. The 
ratio $n_M/n_{INFL}$ remains \cite{mono} essentially 
unaltered after the production of monopoles at the end of 
hybrid inflation until $T_R~$. The initial number density 
of magnetic monopoles can be estimated by the Kibble 
mechanism \cite{kibble} which gives 
$n_M\approx(3p/4\pi)m_{INFL}^3~$, where $p\sim 1/10$ is 
a geometric factor. This, together with 
$n_{INFL}\approx V_{INFL}/m_{INFL}~$, where 
$V_{INFL}=\kappa^2M^4$ is \cite{dss} the constant energy 
density driving hybrid inflation, yields 
$n_M/n_{INFL}\approx 3pm_{INFL}^4/4\pi V_{INFL}$ and, 
thus,
\begin{equation}
\left(\frac{n_M}{s}\right)(T_R)\approx 
\frac{9\sqrt{2}p\kappa T_R}{8\pi M}~\cdot
\label{eq:initial}
\end{equation}
The fraction of the inflationary energy density which goes 
into magnetic monopoles right after the end of 
hybrid inflation is $\approx 3pm_{INFL}^3m_M/V_{INFL}=
3p\kappa m_M/\sqrt{2}\pi M\ll 1$, for all relevant 
$\kappa$'s. Here $m_M\approx 4\pi M/g_G$ is the 
monopole mass, with $g_G$ being the GUT gauge coupling 
constant. The final $n_M/s$ can then be found from 
Eqs.(\ref{eq:dilution}) and (\ref{eq:initial}).

\par
The primordial inflaton decays into a pair of right-handed 
neutrino superfields $\nu^c$ with mass 
$M_{\nu^c}=\eta m_{INFL}/2$ ($0<\eta<1$) via the 
superpotential coupling $\bar{H}^c\bar{H}^cF^cF^c$. The 
decay width is 
$\approx(1/8\pi)(M_{\nu^c}/M)^2(1-\eta^2)^{1/2}
m_{INFL}$. Note that $M_{\nu^c}$ also originates from the 
coupling $\bar{H}^c\bar{H}^cF^cF^c$ and, thus, cannot be 
bigger than about $2M^2/m_P$. The primordial reheat 
temperature is given by
\begin{equation}
T_R\approx
\frac{3^{1/2}5^{1/4}\kappa^{1/2}m_P^{1/2}M_{\nu^c}
(1-\eta^2)^{1/4}}{2\sqrt{2}\pi g_*^{1/4}(T_R)M^{1/2}}~,
\label{eq:primordial}
\end{equation}
where $g_*(T_R)\approx 234.25+75n/2$ since, at $T_R~$, the 
bosonic $D$'s and $H$'s are still massless. 

\par
For each $\kappa$, the value of $M$ can be found by 
simultaneously solving Eqs.(2) and (5) of Ref.\cite{hier} 
with their right hand sides divided by an extra factor of 2 
due to the fact that $H^c$ ($\bar{H}^c$) contains four 
$SU(2)_R$ doublets. We will take, for definiteness, 
$\kappa=4\times 10^{-3}$ which gives 
$M\approx 9.57\times 10^{15}~{\rm{GeV}}$. With these 
values, hybrid inflation ends only when the inflaton field 
$S$ is infinitesimally close to the instability point 
(at $\vert S\vert=M$) of the inflationary trajectory, 
as one can easily deduce from the slow roll conditions. 
Moreover, our present horizon scale crosses outside the 
inflationary horizon at $\vert S\vert\approx 2.63M$. 
From Eq.(\ref{eq:primordial}), we find that $T_R$ can 
take all the values up to about $10^{12}~{\rm{GeV}}$. 
The maximal $T_R$ is obtained at $\eta^2=2/3$ which 
maximizes the decay width of the primordial inflaton. The 
corresponding value of $M_{\nu^c}$ turns out to be much 
smaller than $2M^2/m_P$. It is important to note that the 
stringent gravitino constraint \cite{gravitino} on $T_R$ 
($\stackrel{_{<}}{_{\sim }}10^9~{\rm{GeV}}$) is 
alleviated here since the primordial gravitinos suffer 
considerable dilution during thermal inflation.

\par
We will now estimate the present $D$ and $H$ particle 
abundance. Due to the two unbroken $Z_2$'s under which 
the $D$'s and $H$'s change sign independently, these 
particles can only annihilate in pairs. The annihilation 
processes remain in thermal equilibrium at all temperatures 
between $T_R$ and $T_c~$. Moreover, at temperatures higher 
than $T_c~$, where the PQ (thermal) transition takes place, 
the fermionic $D$'s and $H$'s are massless. On the 
contrary, the scalar components of these fields acquire 
soft SUSY breaking and temperature-dependent masses.  
Consequently, for $m_{3/2}\gg T_c$ (see below), the 
number density of these bosons at temperatures close to 
$T_c$ is suppressed, and we thus ignore them. The 
number density of the fermionic $D$'s ($H$'s), 
for temperatures just above $T_c~$, is  
$n_{D}(T_c)\approx 21n\zeta(3)T_c^3/2\pi^2$ 
($n_{H}(T_c)\approx 14n\zeta(3)T_c^3/2\pi^2$), 
where $\zeta(3)\approx 1.2021$. 

\par
After the completion of thermal transition, the $D$ ($H$) 
superfields acquire intermediate scale masses 
$\sqrt{2}\alpha f_a$ ($\sqrt{2}\beta f_a$). Their 
total relative contribution to the energy density of the 
universe, immediately following the transition, 
is given by $\Omega_{DH}(T_c)\equiv
\rho_{DH}(T_c)/\rho_{infl}(T_c)\approx 7n
(3\alpha+2\beta)\zeta(3)f_aT_c^3/
\sqrt{2}\pi^2V_0\ll 1$,
where $\rho_{infl}$ is the energy density in thermal 
inflatons. $\Omega_{DH}$ remains essentially 
constant until $T_r$ since pair annihilation 
of the fermionic $D$'s and $H$'s is frozen already at 
$T_c$ where these fermions acquire their intermediate 
scale masses. Between $T_r$ and the equidensity temperature 
$T_{eq}\approx 2.5\times 10^{-9}~{\rm{GeV}}$ (for 
present value of the Hubble constant 
$H_0\approx 65~{\rm{km/s~Mpc}}$), where matter and 
radiation have equal energy densities, $\Omega_{DH}$ is 
enhanced by a factor $T_r/T_{eq}$ and remains practically 
unaltered thereafter. So the present abundance of these 
particles is 
\begin{equation}
\Omega_{DH}\approx
\frac{7n(3\alpha+2\beta)\zeta(3)}{\sqrt{2}\pi^2}
~\frac{f_aT_c^3}{V_0}~\frac{T_r}{T_{eq}}~\cdot
\label{eq:abundance}
\end{equation}
This can easily be of order unity so that the $D$ and $H$ 
fermions with intermediate scale masses constitute a 
sizable fraction of the cold dark matter in the present 
universe.

\par
We are now ready to proceed to a numerical example. We 
choose the gravitino mass $m_{3/2}=300~{\rm{GeV}}$ and 
the number of $D$'s and $H$'s $n=7$. The thermal inflaton 
mass is then $m_{infl}\approx 575~{\rm{GeV}}$ and the 
Higgsino mass $\mu\approx 235~{\rm{GeV}}$ (recall that 
we take $\epsilon^2=2/3$). The critical temperature for 
the PQ transition can be evaluated from 
Eq.(\ref{eq:critical}) and turns out to be 
$T_c\approx 69~{\rm{GeV}}$ for coupling constants 
$\alpha=\beta=1$. We further take 
$\lambda_2=2\times 10^{-3}$. From Eqs.(\ref{eq:minimum}) 
and (\ref{eq:energy}), we then obtain the axion decay 
constant $f_a\approx 7.57\times 10^{11}~{\rm{GeV}}$ and 
the vacuum energy density which drives thermal inflation 
$V_0\approx 3.16\times 10^{28}~{\rm{GeV}}^4$. The 
parameter $\lambda_1$ is evaluated from the Higgsino mass
and comes out to be $\approx 10^{-3}$. Eq.(\ref{eq:reheat}) 
then yields $T_r\approx 2.81~{\rm{GeV}}$ for the reheat 
temperature after thermal inflation. 

\par
The present relative monopole number density is estimated from 
Eqs.(\ref{eq:dilution}) and (\ref{eq:initial}). It turns out 
to be $n_M/s\approx 4.6\times 10^{-41}T_R$ for the chosen 
value of $\kappa$ ($=4\times 10^{-3}$). This implies 
that $T_R$'s of order $10^{10}~{\rm{GeV}}$, which can 
be naturally obtained from Eq.(\ref{eq:primordial}) by 
appropriately adjusting $M_{\nu^c}$ (or $\eta$), lead 
to $n_M/s\sim 10^{-30}$. This corresponds to the 
well-known Parker bound for the monopole flux in our 
galaxy derived from galactic magnetic field considerations. 
Needless to say that, by lowering $T_R~$, one can easily 
reduce the predicted monopole flux by a couple of orders 
of magnitude below the Parker bound. However, this flux 
cannot be suppressed much further with natural (not too 
small) values of the coupling constants. In conclusion, 
we have shown that thermal inflation associated with the 
PQ symmetry and the $\mu$ term can naturally suppress 
the present flux of monopoles from SUSY hybrid inflation 
to values below but near the Parker bound. This flux 
should be possibly accessible to ongoing and future 
experiments.

\par
The present abundance of the $D$ and $H$ fermions with masses 
$\approx 1.1\times 10^{12}~{\rm{GeV}}$ is found from
Eq.(\ref{eq:abundance}) to be $\Omega_{DH}\approx 0.185$.
These particles can therefore provide a considerable fraction 
of the cold dark matter. The rest can consist of axions which 
are also present in this scheme. The relic axion abundance 
$\Omega_a$ has been calculated in Ref.\cite{axion}. 
Assuming that the initial value of the axion field is 
about $0.55f_a$, Eq.(13) of the first paper in this 
reference yields $\Omega_a\approx 0.115$ for 
$H_0=65~{\rm{km/s~Mpc}}$ and 
$f_a=7.57\times 10^{11}~{\rm{GeV}}$. We see that one
can naturally obtain a cold dark matter component in the 
universe consisting of both axions and intermediate scale 
mass fermions with total energy density equal to about 
$30\%$ of its critical density, consistent 
with recent observations \cite{cluster,lambda}. It 
should be clear that, by  appropriately choosing the 
values of the parameters, one can easily adjust not only 
the total cold dark matter abundance but also its 
composition of axions and intermediate scale mass fermions.
The present numbers only serve as an example.

\par
As explained, the $D$ and $H$ fermions are stable due 
to the two $Z_2$ remnants of $U(1)_R\times U(1)_{PQ}~$. 
Their stability is absolute if these $U(1)$'s are 
exact symmetries of the superpotential to all orders. 
However, as has been shown \cite{discrete}, global 
$U(1)$'s may be present as effective rather than exact 
symmetries. Indeed, some of the discrete 
symmetries which normally emerge from the underlying 
string theory can effectively behave as continuous 
symmetries. These continuous symmetries are 
expected to be explicitly broken by some higher order 
operators in the superpotential which are allowed by 
the underlying discrete symmetries. If the order of 
these operators is adequately high, they do not 
affect our scheme except that they may provide highly 
suppressed Yukawa couplings for the decay of the 
$D$'s and/or the $H$'s by violating either or both 
the $Z_2$'s. Such couplings could be $DFF$, $DF^cF^c$ 
or $HFF^c$ with coefficients suppressed by 
$(f_a/m_P)^5$ leading to a lifetime 
$\sim 10^{22}~{\rm{years}}$. The long-living $D$'s 
and $H$'s with masses $\sim 10^{12}~{\rm{GeV}}$ may 
then provide an explanation \cite{sdm} of the recently 
observed \cite{uhecr} ultra-high energy cosmic ray 
events. Note that these intermediate scale mass 
particles were introduced for making thermal inflation 
possible, and thus provide a mechanism for diluting 
the monopoles. Their role in dark matter and cosmic 
rays is an extra bonus!

\par
It is generally difficult to generate the baryon 
asymmetry of the universe in models with a low 
reheat temperature such as our scheme. Any pre-existing 
baryon (or lepton) asymmetry is utterly diluted by 
thermal inflation. Moreover, after the subsequent 
reheating, the universe is too cold to allow baryon 
number violation and out-of-equilibrium conditions. 
Baryogenesis mechanisms which may be applicable here 
have been discussed in Refs.\cite{trodden,riotto}. The 
latter uses the fact \cite{turner} that the oscillating 
(thermal) inflaton does not decay instantaneously at 
$T_r$. It rather follows the usual exponential decay 
law. The ``new radiation'', which is so produced, 
reaches a maximum temperature which is much higher than 
the electroweak scale. This radiation then gradually 
cools down and, finally, dominates the energy density 
at $T_r$. During this process, a lepton asymmetry 
can be generated via the Affleck-Dine (AD) mechanism 
\cite{ad}. The decay of the AD condensate is plasma 
blocked at temperatures higher than its frequency of 
oscillations which is expected to be of the order of 
the electroweak scale. Actually, this condensate decays 
at a temperature $T_*\sim M_W$, generating a lepton 
asymmetry of order unity (or smaller), provided that its 
energy density at $T_*$ is comparable to (or smaller 
than) the ``new radiation'' energy density. A fraction 
of this asymmetry is then immediately converted into 
baryon asymmetry by the electroweak sphalerons. From $T_*$ 
until $T_r$, the baryon asymmetry acquires \cite{riotto} 
a dilution factor $(T_r/T_*)^5$ and remains constant 
thereafter. It is clear that this scenario can easily lead 
to an adequate baryogenesis in our scheme.

\par
In summary, we considered the cosmological problem arising 
when hybrid inflation is applied to (SUSY) GUTs which 
predict magnetic monopoles. This problem is due to the 
copious monopole production at the end of inflation. We 
showed that the monopole flux can be naturally reduced to 
values below but near the Parker bound by invoking thermal 
inflation ``driven'' by the PQ symmetry, which also 
generates the $\mu$ term of MSSM. This flux may be 
accessible to ongoing and future experiments. Although 
our mechanism was presented within the SUSY PS model, 
its applicability is much wider. An interesting 
by-product is the presence of intermediate scale mass 
fermions and axions in the cold dark matter of the 
universe. These fermions, which were introduced for 
making thermal inflation possible, may explain the recently 
observed ultra-high energy cosmic rays.  

\vspace{0.5cm}

\par
We thank A. Riotto for useful discussions. This work was 
supported by the European Union under TMR contract number 
ERBFMRX--CT96--0090, the DOE under grant number 
DE-FG02-91ER40626, and the NSF under subcontract PHY-9800748.

\def\ijmp#1#2#3{{ Int. Jour. Mod. Phys. }{\bf #1~}(#2)~#3}
\def\pl#1#2#3{{ Phys. Lett. }{\bf B#1~}(#2)~#3}
\def\zp#1#2#3{{ Z. Phys. }{\bf C#1~}(#2)~#3}
\def\prl#1#2#3{{ Phys. Rev. Lett. }{\bf #1~}(#2)~#3}
\def\rmp#1#2#3{{ Rev. Mod. Phys. }{\bf #1~}(#2)~#3}
\def\prep#1#2#3{{ Phys. Rep. }{\bf #1~}(#2)~#3}
\def\pr#1#2#3{{ Phys. Rev. }{\bf D#1~}(#2)~#3}
\def\np#1#2#3{{ Nucl. Phys. }{\bf B#1~}(#2)~#3}
\def\npps#1#2#3{{ Nucl. Phys. (Proc. Sup.) }{\bf B#1~}(#2)~#3}
\def\mpl#1#2#3{{ Mod. Phys. Lett. }{\bf #1~}(#2)~#3}
\def\arnps#1#2#3{{ Annu. Rev. Nucl. Part. Sci. }{\bf
#1~}(#2)~#3}
\def\sjnp#1#2#3{{ Sov. J. Nucl. Phys. }{\bf #1~}(#2)~#3}
\def\jetp#1#2#3{{ JETP Lett. }{\bf #1~}(#2)~#3}
\def\app#1#2#3{{ Acta Phys. Polon. }{\bf #1~}(#2)~#3}
\def\rnc#1#2#3{{ Riv. Nuovo Cim. }{\bf #1~}(#2)~#3}
\def\ap#1#2#3{{ Ann. Phys. }{\bf #1~}(#2)~#3}
\def\ptp#1#2#3{{ Prog. Theor. Phys. }{\bf #1~}(#2)~#3}
\def\plb#1#2#3{{ Phys. Lett. }{\bf#1B~}(#2)~#3}
\def\apjl#1#2#3{{ Astrophys. J. Lett. }{\bf #1~}(#2)~#3}
\def\n#1#2#3{{ Nature }{\bf #1~}(#2)~#3}
\def\apj#1#2#3{{ Astrophys. J. }{\bf #1~}(#2)~#3}
\def\anj#1#2#3{{ Astron. J. }{\bf #1~}(#2)~#3}
\def\mnras#1#2#3{{ Mon. Not. R. Astron. Soc. }
{\bf #1~}(#2)~#3}
\def\grg#1#2#3{{ Gen. Rel. Grav. }{\bf #1~}(#2)~#3}
\def\s#1#2#3{{ Science }{\bf #1~}(#2)~#3}
\def\baas#1#2#3{{ Bull. Am. Astron. Soc. }{\bf #1~}(#2)~#3}
\def\ibid#1#2#3{{ ibid. }{\bf #1~}(#2)~#3}
\def\cpc#1#2#3{{ Comput. Phys. Commun. }{\bf #1~}(#2)~#3}
\def\astp#1#2#3{{ Astropart. Phys. }{\bf #1~}(#2)~#3}
\def\jpa#1#2#3{{ J. Phys. }{\bf A#1~}(#2)~#3}
\def\pan#1#2#3{{ Phys. Atom. Nucl. }{\bf #1~}(#2)~#3}

\end{document}